\documentclass{elsart}

\usepackage{graphicx}
\usepackage{amssymb}

\begin{document}
\begin{frontmatter}
\title{Slow dynamics, aging, and glassy rheology in soft and living matter}

\author{Ranjini Bandyopadhyay$^{a}$},
\author{Dennis Liang$^{b}$}, 
\author{James L. Harden$^{c}$}
\and
\author{Robert L. Leheny$^{b}$}
\address{$^{a}$Liquid Crystal Laboratory, Raman Research Institute, Bangalore 560080, India,\\
$^{b}$Department of Physics and Astronomy, Johns Hopkins University, Baltimore MD 21218, USA\\
and $^{c}$Department of Physics, University of Ottawa,Ottawa, ON  K1N 6N5,Canada}
\begin{abstract}
We explore the origins of slow dynamics, aging and glassy rheology in soft and living matter. Non-diffusive slow dynamics and aging in materials characterised by crowding of the constituents can be explained in terms of structural rearrangement or remodelling events that occur within the jammed state. In this context, we introduce the jamming phase diagram proposed by Liu and Nagel to understand the ergodic-nonergodic transition in these systems, and discuss recent theoretical attempts to explain the unusual, faster-than-exponential dynamical structure factors observed in jammed soft materials. We next focus on the anomalous rheology (flow and deformation behaviour) ubiquitous in soft matter characterised by metastability and structural disorder, and refer to the Soft Glassy Rheology (SGR) model that quantifies the mechanical response of these systems and predicts aging under suitable conditions. As part of a survey of experimental work related to these issues, we present x-ray photon correlation spectroscopy (XPCS) results of the aging of laponite clay suspensions following rejuvenation. We conclude by exploring the scientific literature for recent theoretical advances in the understanding of these models and for experimental investigations aimed at testing their predictions. 
\end{abstract}

\end{frontmatter}
\section{Introduction}
\label{sec:intro}

Certain out-of-equilibrium soft materials, including foams \cite{mayer}, concentrated emulsions \cite{cip_emulsions} and fractal colloidal gels \cite{cip_colloidalgels}, exhibit slow relaxation processes that often resemble the glassy dynamics observed in disordered hard materials such as polymer melts and spin glasses. The properties of many of these disordered soft systems evolve slowly with time, a behaviour commonly termed `aging'. Aqueous foams, for example, age due to coarsening and bubble rearrangement processes \cite{durian_science}. Polymer glasses   \cite{struik}, microgel solutions \cite{cloitre} and aqueous clay suspensions \cite{knaebel_aging,bellour_aging} are some more examples of aging systems. In spite of their apparent diversity, these materials are all structurally disordered and out-of-equilibrium. The existence of certain universal features in the evolution of the mechanical responses of these systems indicates an underlying generic mechanism that is responsible for the observed behaviour. To this end, a number of theoretical and experimental studies of the  evolution of aging systems -- including calculations of magnetic  
suceptibilities in frustrated spin systems \cite{kurchan}, dielectric measurements of memory in molecular glasses \cite{hasan_epl}, and measurements of the dynamic correlation functions of colloidal gels \cite{cip_colloidalgels} -- have noted commonality in the dynamics under study with seemingly disparate systems.

In addition to being sources of challenging fundamental problems in their own right, soft materials can serve as excellent model systems in the understanding of hard condensed matter as they often display several dynamic and thermodynamic properties typically associated with the latter. For example, recent work on melting of colloidal crystals has elucidated the importance of pre-melting at grain boundaries \cite{yodh_premelting}. Another example is the mode coupling theory (MCT). Originally applied to hard sphere liquids, MCT accounts for many features observed in the colloidal glass transition including the presence of a two-step relaxation in the  
dynamical correlation function \cite{gotze_mct,pusey_dls_glass,weitz_mct}. Based on this success, researchers have sought to extend application of the theory to the glass transition in molecular glasses and polymers \cite{tao,richter}. Owing to their larger sizes and slower relaxation times when compared to, say, atomic fluids, the structure, dynamics and mechanical response of soft materials can be studied experimentally by common laboratory techniques such as microscopy, light scattering and rheology.  
The interaction between the individual constituents of soft materials can be tuned easily \cite{emulsion_manneville} and this feature is exploited for example in the verification of the predictions of the MCT for repulsive and attractive colloidal glasses \cite{vanmegan_hs,bergen_at}. For a more complete discussion, we would like to refer the readers to a recent review article by Cipelletti and Ramos \cite{cip_slow_review}. 

The presence of a glassy response, which is characterized by very slow, history-dependent dynamics, implies a loss of ergodicity. The transition of colloidal suspensions and granular media from an ergodic to a nonergodic state can be understood in terms of the jamming of overcrowded particles, which results in kinetic arrest and ultra-slow relaxation processes \cite{edwards_jam}. 
For dense materials such as glasses, sandpiles and foams, Liu and Nagel \cite{jamming_liu} suggest a 'jamming' phase diagram that considers the possibility of unjamming the system ({\it i.e} driving it towards ergodicity) by increasing its temperature, by applying appropriately large shear stresses, or by decreasing the density of its constituents. Universal non-diffusive slow dynamics and aging have been observed in a variety of jammed soft materials, for example, in colloidal suspensions \cite{cip_emulsions,cip_colloidalgels,knaebel_aging,bellour_aging,segre_gel,ranjini_lapo}, concentrated emulsions \cite{cip_emulsions}, micellar polycrystals \cite{cip_emulsions} and lamellar gels \cite{cip_emulsions,cip_mlv}. Glassy dynamics also occur in biologically relevant systems. Examples include the kinetic arrest and scale-free rheology observed in the cytoskeleton of a living cell \cite{bursac_naturemat,fabry_microrheol} and the ultraslow relaxation processes in models of the adaptive immune response system \cite{sun_immune}. 

In the specific case of disordered soft materials that can flow in response to macroscopic strains, metastability can result in anomalous rheology that is understood in terms of the soft glassy rheology (SGR) theory proposed by Sollich {\it et al.} \cite{sollich_prl,sollich_sgr}. For those systems whose thermal energy is not sufficient to achieve complete structural relaxation, SGR predicts identical power law frequency dependences of the low-frequency elastic and viscous moduli and identifies the exponent of this power law with a mean field noise temperature. The model predicts a glass transition, with the glass phase being characterised by the aging of the viscoelastic moduli and the presence of yield stresses. Many of these features have been observed experimentally in concentrated microgel solutions \cite{ketz}, foams \cite{khan}, paint \cite{mackley}, compressed emulsions \cite{mason_colloid}, a thermotropic liquid crystal with quenched disorder  \cite{ranjini_8CB} and in the cytoskeleton of living cells \cite{fabry_microrheol}.

In the remainder of this article, we discuss some theoretical models that describe the slow dynamics, aging and glassy rheology in soft materials and cite experimental results that verify the predictions of these models. In doing so, we have organised the paper in the following manner: section 2 contains an overview of the jamming phase diagram proposed by Liu and Nagel \cite{jamming_liu} followed by a brief description of theoretical attempts to explain the non-diffusive slow dynamics and aging in soft materials in the jammed state. We next focus on the SGR theory \cite{sollich_prl,sollich_sgr} that quantifies the flow and deformation behaviour of soft glassy materials (SGMs). Section 3 reviews experimental work on the slow dynamics and aging in soft and living matter, followed by observations of anomalous rheology in many of these systems, with some emphasis on our recent work on the aging behaviour in synthetic colloidal laponite clay suspensions \cite{ranjini_lapo} and the observation of soft glassy rheology in a thermotropic smectic liquid crystal confined within the pores of a compliant colloidal gel \cite{ranjini_8CB}. This is followed by a brief summary and concluding remarks in section 4.

\section{Models}
\label{sec:models}
In this section, we present brief overviews of some models that are particularly relevant in the description of the slow dynamics, aging and glassy rheology in soft and living matter. These include the jamming picture, the prediction of ultraslow non-diffusive relaxation processes in aging soft matter and the SGR model. Experiments that verify the predictions of these theories are summed up in section 3.

\subsection{The jamming phase diagram and the compressed exponential dynamical structure factor in some jammed systems}

An approach to understanding the ergodic-nonergodic transition is provided by the concept of jamming \cite{edwards_jam}. Jamming of overcrowded particles results in kinetic arrest and signals a fluid-solid transition. The jamming phase diagram provides a unified picture for the ergodic-nonergodic transition in systems as diverse as molecular glasses, foams, granular matter and colloidal suspensions  \cite{jamming_liu}. If a system arrests as soon as it is able to support an external load, such that it responds elastically to some compatible loads while undergoing plastic rearrangments under incompatible loads, it is classified as 'fragile' \cite{farr_epl,farr_pre,cates_jam}. Jamming of fragile matter leads to slow dynamics because of the kinetic constraints it imposes on the motion of the individual constituents. Cates {\em et al.} draw a distinction between fragile materials and conventional solids by noting that in the case of the former, applied stresses are supported by force chains that can collapse under the action of an incompatible load \cite{cates_jam,cates_review}. This can lead to unjamming. The jamming picture has been subsequently expanded by Liu and Nagel \cite{jamming_liu} to include jammed systems such as foams and granular matter which are characterised by yield stresses and are therefore not fragile according to the definition of Cates {\em et al.} \cite{cates_jam}. In addition to fluidising a system by the application of shear stresses as suggested in \cite{cates_jam}, the authors in \cite{jamming_liu} suggest the possibility of unjamming a jammed system by increasing its temperature (in the case of, say, molecular glasses) or by decreasing the density of its constituents (in the case of, say, aqueous foams).  In this way, they suggest a natural connection between the glass transition in supercooled liquids and the fluid-solid transition in athermal systems such as foams and granular matter. To  
encompass these ingredients, the authors introduce a phase diagram (Fig. 1a) for jammed systems whose axes depend on temperature, load and the inverse of the density, such that appropriate changes in these parameters can result in the onset of the unjamming transition. 

\begin{figure}[ht!]
\centering
\includegraphics[width=12.cm]{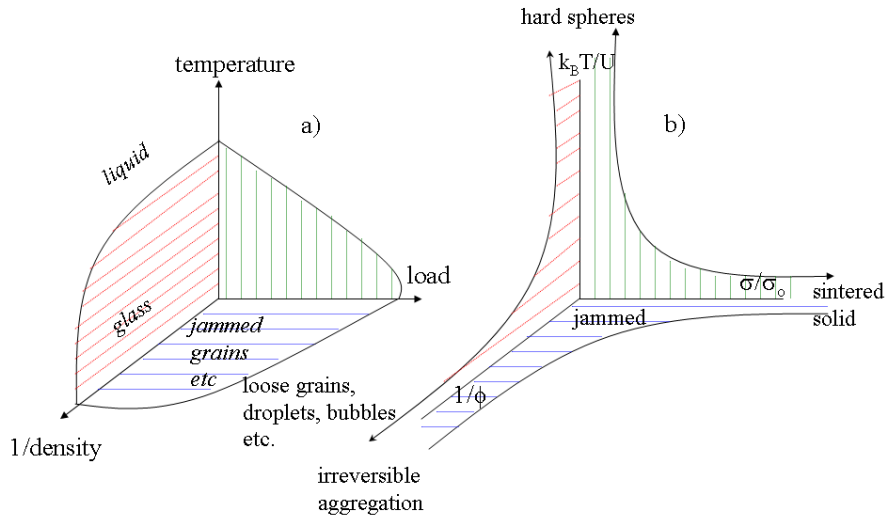}
\caption{\label{p1246} The jamming phase diagram proposed in \cite{jamming_liu} is depicted in a). The jamming phase diagram for attractive colloidal particles constructed on the basis of experiments reported in \cite{trappe_jamming} is depicted in b), where $\phi$ is the colloidal volume fraction, $U$ is the interparticle attractive energy whose scale is set by the thermal energy $k_{B}$T and $\sigma/\sigma_{\circ}$ is a suitably normalised applied stress. The shaded regions in both diagrams indicate jamming. The phase boundaries are qualitative, and may vary between systems. Figures adapted from \cite{jamming_liu} and \cite{trappe_jamming}.}
\end{figure}

Ultra-slow dynamics and aging observed in colloidal gels \cite{cip_colloidalgels}, clay suspensions \cite{knaebel_aging,bellour_aging,ranjini_lapo}, compressed emulsions \cite{cip_emulsions} and multilamellar vesicles \cite{cip_mlv} are understood in terms of the relaxation of internal stresses that are built into the sample at the jamming transition. The dynamical structure factors (DSFs) characterising these systems are strongly dependent on sample age (quantified in terms of the waiting time $t_{a}$ since sample preparation or the onset of jamming) and exhibit non-diffusive, faster-than-exponential forms. The relaxation time $\tau_{s}$ characterising these slow dynamics scales with wave vector $q$ as $\tau_{s} \sim q^{-1}$. In the case of a fractal colloidal gel which undergoes syneresis (microcollapses in its structure) \cite{cip_emulsions,cip_colloidalgels}, the relaxation process that gives rise to the observed decay of the DSF is modeled in terms of the elastic response to local stress
sources that appear at random in the medium. Each stress source is approximated as a force dipole, such that for a small number density of stress sources distributed randomly in space, calculations predict the relation $\tau_{s} \sim q^{-1}$ and a compressed exponential form for the DSF: $f(q,t) \sim \exp[-(qt)^{p}]$ with a compressing exponent $p \thickapprox$ 1.5. These trends imply ballistic strain in the sample characterised by a velocity distribution function $W(v) \sim v^{-(p+1)}$ in response to the stress \cite{cip_emulsions}. Allowing for a finite collapse time $\theta$ for each microcollapse event and writing a central force elasticity equation for the strain field around a force dipole, Bouchaud and Pitard \cite{bouchaud_dls} calculated several regimes for the DSF, with the compressed exponential behaviour $f(q,t) \sim \exp[-(qt)^{1.5}]$ occuring only at earlier times. With a coupling between stress dipoles incorporated into the model, the growth of $\tau_s$ with $t_a$ is also computed, giving a $\tau_s \sim t_a^{2/3}$ dependence in this early time regime. The compressed exponential form of the DSF and the inverse linear dependence of the relaxation time on the wave vector are seen in experiments in a wide variety of soft materials. A sublinear dependence of $\tau_{s}$ on $t_{a}$ is also often observed, though the exponent of the power law can deviate significantly from the theoretical prediction in \cite{bouchaud_dls}. This variability indicates a strong dependence of the aging process on the structural and dynamical details of the sample under investigation.

\subsection{Trap models and Soft Glassy rheology (SGR)}
Two-step relaxation processes, such as those observed in supercooled liquids \cite{tao,richter}, dense colloidal suspensions \cite{pusey_dls_glass} and concentrated emulsions \cite{krall1}, can be understood in terms of the trap model.  This model describes the glassy dynamics of a particle (or an `element') trapped in a cage, {\em i.e.} in a potential well created by the constraints imposed by neighbouring particles, such that escape from a cage is possible only by some activated process (for example, by thermal activation in the case of supercooled liquids). These potential energy minima represent metastable states that are surrounded by high energy barriers whose heights, in the case of supercooled liquids, increase with decreasing temperature. At short times, the particles diffuse in their cages while at longer times they escape through activation. The separation of time scales for these processes leads to a two-step relaxation. In the glassy state, the escape process becomes inaccessibly slow leading to nonergodic, partial relaxation and kinetic arrest. In polyelectrolyte microgel pastes, for example, individual microgel beads are trapped in cages formed by the neighboring particles. The intensity autocorrelation functions measured for suspensions of these particles, which can execute sub-diffusive motion in their cages at short times and are trapped (kinetically arrested) at long times, display an incomplete decay \cite{cloitre_paste2}. Bouchaud used the trap model to understand the relation between aging and weak ergodicity breaking in spin glasses \cite{bouchaud_aging}, where an effective particle is trapped in a one-dimensional potential energy landscape, and which, when activated, can hop into another trap. Detailed calculations by Monthus and Bouchaud \cite{bouchaud_trap} show a dynamical phase transition between an ergodic (liquid) phase at high temperature and a non-ergodic aging (glass) phase at low temperature when the density of energy barriers decays exponentially. 

\begin{figure}[ht1!]
\centering
\includegraphics[width=10.cm]{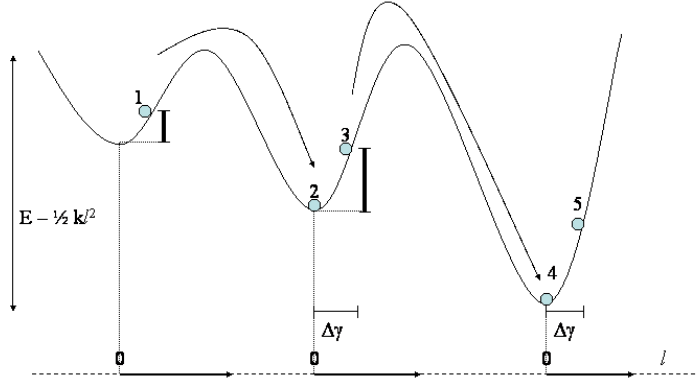}
\caption{\label{p1247} The potential well picture that forms the basis of the SGR theory for metastable and disordered soft systems. The potential energy landscape is characterised by a distribution of yield energies E and local strains $l$. As the system is sheared beyond a yield point (described by a finite value for the yield strain $l_{y}$), stress relaxation occurs by the rearrangement of particles (or `elements') and the system settles to a new local equilibrium configuration. The different configurations of the system are shown as circles numbered from 1 to 5. $\Delta\gamma$ corresponds to displacements when the strain is less than the yield value necessary for the system to hop into the next potential energy minimum. When the yield strain is large enough, the particle can undergo noise-induced hops into the next minimum, and the energy dissipated in such activated yielding processes (denoted by curved arrows) is indicated by the solid vertical bars. Figure adapted from \cite{sollich_sgr}.}
\end{figure}

Sollich {\em et al.} \cite{sollich_prl,sollich_sgr} generalised the trap model to understand the low-frequency rheology of SGMs. The SGR model proposed by Sollich {\em et al.}  \cite{sollich_prl,sollich_sgr} describes the flow and deformation behaviour of soft materials such as colloidal pastes, foams, emulsions and slurries that are characterised by metastability and disorder. The `glassiness' of these materials is a consequence of their structural disorder and metastability, while their `softness' arises from their ability to flow upon the application of macroscopic strains. SGR \cite{sollich_prl,sollich_sgr} uses a generalised trap model to describe SGMs, such that each individual element, which moves in a one-dimensional piecewise quadratic potential (Fig. 2), can undergo `activated yielding' processes due to structural rearrangement events in the system. Such structural rearrangement events are strongly coupled and the resulting interactions are quantified in terms of an effective noise temperature $x$. Exact rheological constitutive equations are derived within the framework of the SGR model, with the dynamic shear modulus showing a power law frequency dependence at low frequencies; for an applied oscillatory strain $\gamma (t) = \gamma \cos \omega t$, the elastic modulus $G^{\prime} \sim \omega^{x-1}$ for $1<x<2$, while the viscous modulus $G^{\prime\prime} \sim \omega^{x-1}$ for $1<x<3$. The viscous and elastic moduli become increasingly frequency independent as $x \to$ 1, with the ratio $\frac{G^{\prime\prime}}{G^{\prime}}$ approaching $(x-1)$ as the glass transition (identified at $x = x_{g} =$ 1) is approached. For $x<x_{g}$, no equilibrium state exists, implying weak ergodicity breaking and the presence of aging phenomena. The flow curve is then characterised by an yield stress $\sigma_{y}$ and exhibits the Herschel-Bulkley form: $\sigma - \sigma_{y} \sim \dot\gamma^{1-x}$. This model predicts aging at $x<1$ as long as $\sigma_{y}$ is not exceeded. For appropriately high values of the noise temperature $x$, Maxwellian relaxation behaviour (characterised by $G^{\prime} \sim \omega^{2}$ at $x >$ 3 and $G^{\prime\prime} \sim \omega$ for $x >$ 2) and a power law flow curve ($\sigma \sim \dot\gamma^{x-1}$ for $x >$ 1) are recovered. 

There have been numerous other attempts to describe the rheology of SGMs using such theoretical approaches. For example, Berthier formulated a microscopic non-equilibrium MCT that describes the yield stresses, flow heterogeneities and activated processes in an SGM that is characterised by a strong coupling between its slow dynamics and the shearing forces \cite{berthier}. The nonlinear rheological behaviour of SGMs characterised by a power law flow curve and the existence of a finite yield stress has also been treated within the framework of the mode coupling model by H\'ebraud and Lequeux \cite{hebraud_mct}. Lequeux and Ajdari \cite{ajdari_sgr} calculate a Vogel-Fulcher type  divergence for the viscosity  in their model of SGMs. Calculations by Head {\em et al.} indicate that when the noise temperature $x$ of the SGM is a decreasing function of the global stress $\sigma$, the flow curve is non-monotonic and the model displays hysteresis \cite{head_sgr}. In the presence of diffusion, calculations based on hopping models for glassy systems \cite{evans_sgr} show strong violations of the fluctuation-dissipation theorem (FDT) when applied to the self-diffusion of small probes in the SGM, while agreement with the FDT is restored when the probes are large enough that the surrounding medium can be viewed as a continuum. It should be noted that these models do not consider the spatial and temporal correlations of the shear-induced rearrangement events that are likely to be present in real systems such as dense emulsions \cite{dense_emulsions_hebraud}. The FDT is found to be violated for aging and driven systems. However, using a non-equilibrium version of the FD relation, Sollich and Fielding show the equivalence between a stationary, weakly driven system and an infinitely aging one \cite{field_fdr}.  

\section{Soft glassy materials: review of experiments}
\label{sec:expm}

In this section, we discuss experimental studies related to the predictions of the models introduced in the previous section. Most of these experimental results are obtained using one of two experimental techniques, {\em viz.} rheology and photon correlation spectroscopy (PCS), the principles of which we will first outline in the following subsection. 

\subsection{Experimental techniques}

PCS experiments \cite{pecora} in the single scattering limit involve the measurement of the wave vector dependent intensity autocorrelation function $g_{2}(q,t) = \frac{<I(q,t^{\prime})I(q,t^{\prime}+t)>_{T}}{{<I(q,t^{\prime})>_{T}^{2}}}$, where $I(q,t^{\prime})$ is the scattered intensity  measured at time $t^{\prime}$, $t$ is a delay time and $T$ is the total averaging time. The measured $g_{2}(q,t)$ is related to the DSF $f(q,t)$ {\em via} the Siegert relation $g_{2}(q,t) = 1 + \beta {|{f(q,t)}|^{2}}$ where $\beta$, the coherence factor, depends on the experimental optics and $f(q,t)$ contains information about the dynamics of the scatterers. For example, for a dilute suspension of monodisperse colloidal beads with purely diffusive dynamics, $f(q,t) \sim \exp(\frac{-t}{\tau_{\circ}})$ , where $\tau_{\circ}$ is the characteristic time scale for scatterer diffusion and is related to the self-diffusion coefficient $D_{\circ}$ according to the formula $\tau_{\circ} = (D_{\circ}q^{2})^{-1}$. X-ray photon correlation spectroscopy (XPCS) uses synchrotron radiation to probe very high values of $q$ that are inaccessible by PCS methods involving visible light \cite{xpcs1,xpcs2}. For example, XPCS has been employed in \cite{ranjini_lapo} to study the particle scale dynamics in an aging clay suspension, where each particle is a disc approximately 15 $nm$ in radius and 1 $nm$ in thickness. The dynamics of multiply scattering samples can be accessed with diffusing wave spectroscopy (DWS) \cite{dws}. Multispeckle DWS measurements of probe particles diffusing in aging clay suspensions are used to characterise the aging behaviour of the medium and have been reported in \cite{knaebel_aging}. Conventional PCS experiments use a point detector and therefore require considerable time-averaging of the intensity autocorrelation function to ensure reasonable statistics. In order to reduce total experimental time, multi-element detectors, such as charge-coupled device (CCD) cameras, are used to autocorrelate intensities at every camera pixel. For isotropic, nonergodic materials in which a time average and ensemble average are not equivalent, such multispeckle spectroscopy \cite{mdls} provides a key advantage in that the ensemble average may  
be obtained by an average over pixels corresponding to the same wave vector.  For these reasons, multispeckle spectroscopy has proved to be an excellent tool to probe materials with slow and spatially  
heterogeneous dynamics \cite{cip_emulsions,cip_colloidalgels,knaebel_aging,ranjini_lapo}.

In contrast to PCS which measures microscopic properties, rheological experiments quantify the bulk flow and deformation behaviour of materials. Bulk rheology involves the measurements of the viscoelastic moduli and the nonlinear flow properties of materials \cite{macosko}. Depending on the magnitude of the applied strain, rheological measurements are classified as linear and nonlinear. Linear rheology measurements cause no appreciable change in the  
microstructure of the sheared sample, and for a system in equilibrium, the resulting linear response functions can be related to dynamic correlation functions for the system through fluctuation-dissipation relations. For linear measurements involving the application of an oscillatory strain with an angular frequency $\omega$ and a suitably low  
amplitude $\gamma$, measurements provide the resultant stress amplitude $\sigma(\omega)$ and the angle $\delta(\omega)$ by which the measured stress is phase-shifted with respect to the applied strain. The elastic modulus $G^{\prime}$ and the viscous modulus $G^{\prime\prime}$ are then computed from this data using the relations $G^{\prime}(\omega) = {\frac{\sigma(\omega)}{\gamma}}\cos\delta(\omega)$ and $G^{\prime\prime}(\omega) = {\frac{\sigma(\omega)}{\gamma}}\sin\delta(\omega)$. For a conventional viscoelastic material such as a Maxwell fluid, bulk linear rheology provides material parameters such as the relaxation time $\tau_{R}$ and the high frequency plateau modulus $G_{\circ}$ \cite{cates_maxwell}. Large values of applied strains, by contrast, can result in significant changes in the microscopic structure of the material and can give rise to highly nonlinear effects such as shear thinning and thickening, nonzero normal stresses {\em etc.} \cite{macosko}. Typical nonlinear rheology experiments include the measurement of the flow curve (the stress $\sigma$ {\em vs.} the shear rate $\dot\gamma$ plot that provides information on the shear rate dependent viscosity $\eta(\dot\gamma)$) and yield stresses of materials (the stress beyond which the sample begins flowing).

\subsection{Slow dynamics and aging in some jammed systems}

In their work on jammed fragile systems, Cates {\em et al.} consider hard spheres large enough that their thermal motion can be neglected \cite{cates_jam}. Liu and Nagel \cite{jamming_liu}, while proposing the jamming phase diagram  for supercooled liquids, closely-packed grains and bubbles, speculate on the role of interparticle interactions in the jamming phenomenon. These predictions have been verified experimentally by Trappe {\em et al.} who construct a jamming phase diagram (adapted in Fig. 1b) for attractive colloidal particles \cite{trappe_jamming}. The onset of jamming in these experiments is identified with the appearance of a low-frequency plateau of the elastic modulus that signifies the formation of an interconnected solid network \cite{trappe_carbon} that is reminiscent of the stress-supporting force chains suggested in \cite{cates_jam}. Owing to the attractive interparticle interactions that restrict the mobility of the individual constituents, jamming is found to set in at smaller values of the particle volume fraction $\phi$ than in the case of hard spheres. The jamming-unjamming transition in this system is therefore strongly dependent on the interparticle interaction strength $U$. Like in the case of hard spheres, jammed attractive particles can also be unjammed by the application of stresses or by decreasing $\phi$. The composite phase diagram proposed in \cite{trappe_jamming} has $\phi^{-1}$, a suitably normalised applied stress $\sigma/\sigma_{\circ}$ ($\sigma_{\circ} = k_{B}T/a^{3}$, where $a$ is the radius of the colloidal particle) and $k_{B}T/U$ as the three axes, such that an appropriate increase in one or more of these parameters results in fluidisation/ unjamming. 

A feature of systems near the jamming transition, such as supercooled liquids and glasses, is the presence of dynamical heterogeneities \cite{cip_slow_review,richert}. The slow and heterogeneous dynamics in jammed systems such as sand \cite{maj_nature}, colloidal hard sphere glasses \cite{weeks_science} and coarsening foams \cite{mayer_foam} have been studied in detail.  Due to the broad distribution in the depths of the energy barriers in the potential energy landscape characterising these systems, the mechanical properties of SGMs evolve or `age' continuously. Soft colloidal pastes formed from jammed microgel particles show many history-dependent properties that are interpreted in terms of aging phenomena \cite{cloitre,cloitre_paste2}. Their strain recovery depends on the wait time (age) $t_{a}$ after flow cessation and can be scaled onto a master curve for all applied stresses below the yield stress $\sigma_{y}$. This type of scaling is characteristic of aging samples. This aging behaviour can be interrupted by applying large enough shears, a process called rejuvenation. Perhaps the most important result of this work is the observation that the amplitude of the stress uniquely determines the long-time memory and slow evolution of the colloidal paste. Microrheological measurements of the time-dependent creep compliance of an aging cytoskeleton \cite{bursac_naturemat}, which essentially comprises a crowded network of semi-flexible biopolymers, can be scaled onto a master curve very similar to the kind observed for microgel pastes \cite{cloitre}. The aging behaviour, intermittency and slow dynamics observed in this living system strongly resemble the dynamics observed in inert soft glasses.

\begin{figure}[ht2!]
\centering
\includegraphics[width=10.cm]{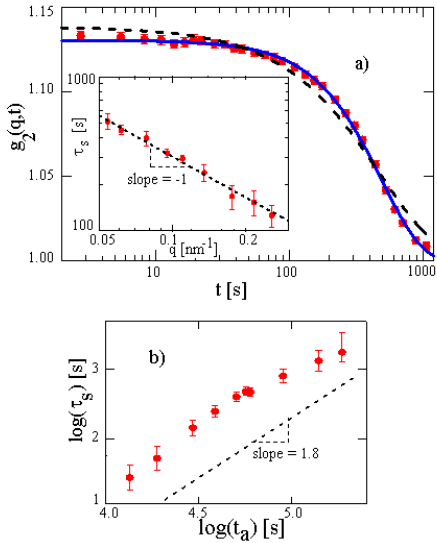}
\caption{\label{p1248} XPCS data (denoted by circles) for the intensity correlation function $g_{2}(q,t)$ for 3.0 wt.\% laponite at $t_{a}$ = 90000s measured at $q$ = 0.135 $nm^{-1}$ is shown in a). The solid line is the result of a fit to the form $g_{2}(q,t)= 1 + \beta[\exp(-(t/\tau_{s})^{1.5})]^{2}$, where $\beta = bA^{2}$ depends on the Siegert factor $b$ and the amplitude of the short-time motion through $A$ as discussed in detail in \cite{ranjini_lapo}. The dashed line in a) shows a poor fit of the data to a simple exponential and is meant to highlight the compressed exponential nature of the observed decay of $g_{2}(q,t)$. The $q$ dependence of the relaxation time $\tau_{s}$ at $t_{a}$ = 40000s is described by the relation $\tau_{s} \sim q^{-1}$ and is shown in the inset of a). Part b) shows a plot of $\log(\tau_{s})$ {\em vs.} $\log$(t$_{a})$ at $q = 0.135 nm^{-1}$. Superlinear power law aging ($\tau_{s} \sim t_{a}^{1.8}$) over the entire range of accesible $t_{a}$s is observed. Some of the data for this figure has been adapted from \cite{ranjini_lapo}.}
\end{figure}

Soft materials at the jamming transition often exhibit aging phenomena and universal, non-diffusive relaxation processes
\cite{cip_emulsions,cip_slow_review}. Interestingly, for systems as diverse as fractal colloidal gels, concentrated emulsions, micellar polycrystals, lamellar gels and synthetic clay suspensions \cite{cip_emulsions,cip_colloidalgels,knaebel_aging,bellour_aging,ranjini_lapo}, the DSF $f(q,t)$ measured by PCS exhibits a two step decay. In spite of the arrested dynamics expected in these systems, $f(q,t)$ is found to
decay completely. In some cases, the characteristic relaxation time $\tau_{f}$ measured for the faster, initial
decay scales with  $q$ as $\tau_{f} \sim q^{-2}$, indicating caged diffusive dynamics \cite{cip_emulsions}.  In the case of a fractal colloidal gel, the DSF $f(q,t)$ characterising the initial decay exhibits
a stretched exponential form with an exponent of 0.7 \cite{krall} and is understood in terms of the contribution of thermally excited, overdamped modes of the gel strands. The second, slower relaxation in these soft glassy materials is more counter-intuitive and is characterised by a DSF with a compressed exponential form ($f(q,t) \sim \exp[-(t/\tau_{s})^{p}]$, where $p \approx$ 1.5), with $\tau_{s}$ varying with $q$ according to the relation $\tau_{s} \sim q^{-1}$. As an example, Fig. 3a shows the intensity autocorrelation function $g_{2}(q,t)$ (indicated by circles)
measured at $q$ = 0.135 $nm^{-1}$ and $t_{a}$ = 90000 s for a 3 wt.\% laponite clay suspension \cite{ranjini_lapo}. The solid line shows a fit to $g_{2}(q,t)$ corresponding to the following form for $f(q,t)$:  $f(q,t) \sim \exp[-(t/\tau_{s})^{1.5}]$ \cite{ranjini_lapo}. This compressed exponential form of the DSF describes the data remarkably well when compared to a fit of the same data to a simple exponential form (indicated by the dashed line). The characteristic time scales $\tau_{s}$ when plotted {\em vs.} $q$ follow the relation $\tau_{s} \sim q^{-1}$ at every
sample age (inset of Fig. 3a shows the data at $t_{a}$ = 40000s).  Evidence of such non-diffusive slow relaxation processes in aging laponite suspensions, but at the much smaller wave vectors accessible by DLS, have also been reported in \cite{bellour_aging}.

In sharp contrast to the slow relaxation processes in supercooled liquids whose DSFs are characterised by stretched exponential forms $f(q,t) \sim \exp[-(t/\tau_{s})^p]$, where $p <$ 1 and $\tau \sim 1/q^2$ (with corrections for de Gennes narrowing \cite{tolle}), the observed faster-than-exponential relaxation of these soft glassy materials is inconsistent with diffusive relaxation processes. Instead, as described in Section 2.1, this unusual form of the DSF and the inverse linear dependence of $\tau_{s}$ on $q$ are understood in terms the ballistic motion of elastic strain deformations that occurs in response to slowly developing localized sites of stresses \cite{cip_colloidalgels,bouchaud_dls}.  Indeed, the picture developed heuristically by Cipelletti {\it et al.} \cite{cip_emulsions} and developed by Bouchaud and Pitard \cite{bouchaud_dls}, in which these sites are identified with randomly positioned stress dipoles whose intensity grows linearly in time, accounts accurately for the form of the DSF and its wave vector dependence.  However, the large array of materials in which these dynamics have observed experimentally \cite{cip_emulsions,cip_colloidalgels,knaebel_aging,bellour_aging,ranjini_lapo,brian_depletion} points to a more general principle underlying the formation of such sites of stress in jammed systems and calls for further investigation, both theoretically and experimentally, into this phenomenon.

\begin{figure}[ht3!]
\centering
\includegraphics[width=8.cm]{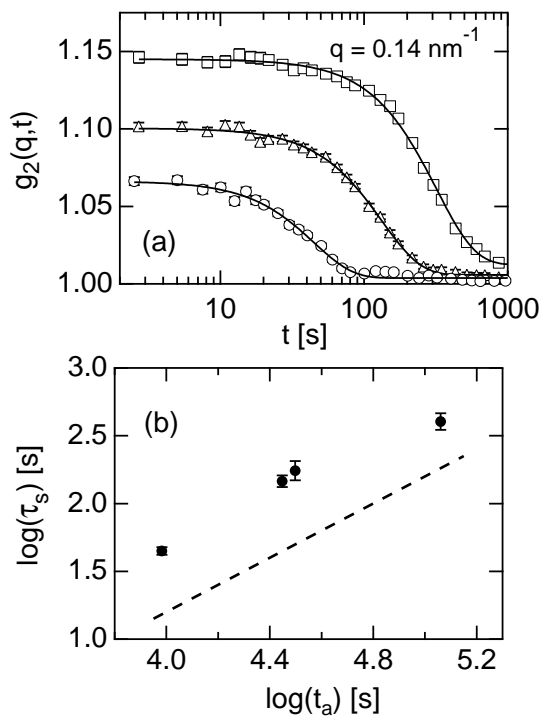}
\caption{\label{p1249} (a) Intensity autocorrelation function $g_{2}(q,t)$ at $q=0.14$ nm$^{-1}$ for a 3 wt.~\% laponite suspension initially aged for 3 days and then rejuvenated by filtering. Results are shown for three times since rejuvenation:  $t_a$ = 9600 s (circles), 28000 s (triangles) and 115000 s (squares).  Solid lines are the results of fits to a compressed exponential lineshape with compression exponent 1.5.  (b) Circles indicate the characteristic relaxation times extracted from $g_{2}(q,t)$ at $q = 0.14$ nm$^{-1}$ as a function of time since rejuvenation.  The dashed line has a slope of 1.0, indicating a linear relationship, $\tau_s \sim t_a$.}
\end{figure}

A common feature of these non-diffusive slow dynamics is the steady evolution, or aging, of the characteristic relaxation time $\tau_{s}$.  However, in contrast to the compressed exponential shape of $f(q,t)$, which is remarkably universal, the precise nature of the aging  behavior is strongly system dependent.  For instance, $\tau_{s}$ varies as a power law with system age $t_{a}$ for colloidal fractal gels, micellar polycrystals, and lamellar gels, with power-law exponents in the range between 0.4 and 0.9 \cite{cip_emulsions}, in rough agreement with the prediction of the Bouchaud-Pitard model, $\tau_{s} \sim t_{a}^{2/3}$.   However, experiments on compressed emulsions \cite{cip_emulsions} show a superlinear dependence of $\tau_{s}$ on $t_{a}$, in sharp contrast to the predicted behavior. As shown in Fig.~3b, the XPCS studies on laponite clay suspensions  \cite{ranjini_lapo} similarly display superlinear behavior with $\tau_{s} \sim t_{a}^{1.8}$, where $t_a$ is the time since formation of the suspensions.  DWS and XPCS measurements of aging depletion gels formed from silica nanoparticles also exhibit compressed exponential  
relaxation processes that evolve with time after the cessation of strong shear as $\tau_{s} \sim t_{a}$ \cite{brian_depletion}. In dynamic light scattering (DLS) studies of laponite and colloidal gels, an initial exponential increase of $\tau_{s}$ with $t_{a}$ is observed, followed by a regime of `full aging' where $\tau_{s}$  
varies as a power-law with $t_{a}$ \cite{cip_colloidalgels,knaebel_aging,bellour_aging}. Thus, no clear pattern has emerged for the temporal evolution of these non-diffusive dynamics.

The sensitivity of the aging behavior of these slow dynamics to details of the system is further illustrated by recent XPCS measurements by our group on laponite suspensions following rejuvenation.  Rejuvenation was accomplished by subjecting suspensions that have aged for long periods (3 days in this case) to strong shear flow by filtering them through 11 $\mu$m diameter pores.  The suspensions initially re-fluidise due to the filtering, after which aging resumes and they progressively regain their solid-like consistency.  Fig.~4a displays $g_2(q,t)$ at $q=0.14$ nm$^{-1}$ for a 3 wt.\% laponite suspension measured at three times since rejuvenation. The solid lines are results of fits to the compressed exponential form with exponent 1.5, identical to the line shape observed for laponite suspensions during aging after their initial formation, as shown in Fig.~3a.  The characteristic relaxation times $\tau_{s}$ at $q = 0.14$ nm$^{-1}$ extracted at various times since rejuvenation $t_{a}$ are shown in Fig.~4b. The dashed line in the figure has slope 1.0, indicating $\tau_s \sim t_a$, clearly distinct from the XPCS results for initially formed laponite, $\tau_s \sim t_a^{1.8}$, shown in Fig.~3b \cite{ranjini_lapo}.  We attribute this contrasting aging behaviour to differing origins of local stress in each case.  As described in \cite{ranjini_lapo}, sites of developing stress in the initially formed suspensions are identified with a growing interparticle repulsion due to charge dissociation.  In the rejuvenation, on the other hand, we identify them with residual local stress that is loaded into the sample by the strong shear flow and that is relieved by slow particle rearrangments.  This more heterogeneous stress distribution, while leading to the same distribution of strain velocities and hence to the same form of $f(q,t)$, evidently evolves differently in time, creating distinct aging behavior.

We note, however, that these new XPCS results on laponite rejuvenation suggest a need to re-evaluate the role of growing interparticle repulsions in the formation of soft solid laponite suspensions. As described in \cite{ranjini_lapo}, a key piece of evidence for such growth was a steady increase with age of the short-time plateau value of $g_2(q,t)$ and the increasingly constrained caged particle motion inferred from this increase.  As Fig.~4a illustrates, laponite suspensions following rejuvenation unexpectedly display a very similar increase in the plateau value despite the fact that the interparticle interactions are likely unaffected by the rejuvenation.  One possibility for this increase in the rejuvenated laponite suspensions is a slow strengthening of force networks in the suspensions weakened by the shear flow and a concomitant decrease in thermal fluctuations of these networks that create a fast
partial decay in $g_2(q,t)$ in analogy with the picture developed in~\cite{krall}.  Thus, the increase in the plateau value during formation and after rejuvenation could have different origins.  However, the strong similarity of the two behaviors suggests otherwise, calling into question the interpretation introduced in \cite{ranjini_lapo}.  Further study of laponite rejuvenation with XPCS would be helpful in resolving this issue. Nevertheless, the contrast between Figs.~3b and 4b illustrates clearly how the aging behavior of the slow, non-diffusive dynamics is sensitive to the
precise source of the local stress.

\subsection{The SGR model: experimental status}

Glassy rheology is a common feature in soft materials characterised by metastability and disorder. A signature feature of soft glassy rheology is the weak power law behaviours of the viscoelastic moduli. Experimental systems that verify the power law behaviour of soft glassy materials include concentrated microgel solutions \cite{ketz}, foams \cite{khan},
paint \cite{mackley} and compressed emulsions \cite{mason_colloid}, all of which have $G^{\prime}$s that show a weak power law frequency dependence, with the exponent ($x$-1) of the power law lying in the range 0.1-0.3. That is, the noise temperature $x$ within the SGR model lies in the range 1.1-1.3.  In the case of the compressed emulsions \cite{mason_colloid}, $G^{\prime}$ is always significantly greater than $G^{\prime\prime}$ above the random close packing fraction ($\phi_{rcp}$ $\sim$ 0.635) of the droplets. As the droplet volume fraction $\phi$ is increased beyond
$\phi_{rcp}$, the droplets squeeze against each other and deform, and the resulting constraints on the droplet dynamics manifest as an increasingly flat elastic response.  The elastic moduli of hard sphere colloidal suspensions, over a limited range of frequencies, also show power law frequency dependences characterised by exponents that decrease
monotonically as the colloidal volume fraction $\phi$ increases toward the kinetic glass transition $\phi_{g}=0.58$ \cite{mason_glass}.  These results are understood in terms of the kinetic arrest of particles trapped in their cages as the kinetic glass transition is approached.  This decrease in the power law exponent towards $(x-1)=0$  indicates an approach towards the glass transition of the SGR model and thus directly connects that transition, at least qualitatively, to the kinetic glass transition in hard sphere colloids.  Further, for volume fractions above $\phi_{g}$,
colloidal suspensions display aging~\cite{courtland}.

Creep experiments on onion phases (20\% AOT in a 15 g/l brine solution) also uncover very slow relaxation processes \cite{roux_langmuir}. The frequency dependent moduli, which track one another over more than three decades of angular frequency, show two distinct regimes: a low-frequency power law regime with an exponent close to zero that describes the slow relaxation of grain boundaries and long length-scale reorganisation processes, and a high-frequency regime which is dominated by the response of individual onions. The first part of the response is strongly reminiscent of SGR. The elastic and viscous moduli of suspensions of aging laponite, after showing Maxwellian behaviour at the earliest times, eventually exhibit power law frequency dependences, with the exponent of the power law becoming smaller as the sample ages \cite{lapo_sgr}.  This observation of aging in laponite when exponent $(x-1)>0$ at first glance seems at odds with the SGR picture, since $x>1$ constitutes the ergodic regime in the model.  However, as mentioned above, studies of laponite indicate that the aging is driven at least partially by an evolving interparticle potential ~\cite{ranjini_lapo}, a feature that clearly lies outside the range of the model.

Another system exhibiting signature features of soft glassy rheology is the thermotropic smectic liquid crystal octylcyanobiphenyl (8CB) confined within the pores of a highly porous and compliant hydrophilic aerosil gel (size of individual aerosil particles is 7 nm) ~\cite{ranjini_8CB}. In this system $G'$ and $G''$ show power law frequency dependencies that arise from an interplay of the smectic order and the quenched disorder introduced by the gel. For example, Fig.~5a displays $G'(\omega)$ at 21$^{\circ}$C (red circles), several degrees below the nematic to smectic transition, and at 36$^{\circ}$C (blue squares), above the transition.  At high temperature, $G'$ is dominated by the colloidal gel. The greatly enhanced modulus in the smectic phase at lower temperature includes a weak power law component whose exponent $(x-1)$ decreases with increasing aerosil density $\rho_s$, as shown in the inset to Fig.~5a.  This decrease thus signals an approach to the glass transition of the SGR model, $x = 1$,~\cite{sollich_prl,sollich_sgr}
with increasing quenched disorder (quantified by $\rho_s$).  The key role of the smectic order in this rheology is illustrated by contrasting this behavior with the rheology of the simple organic liquid dibutyl phthalate confined to the pores of an aerosil gel, shown in Fig 5b.  $G'(\omega)$ measured at  21$^{\circ}$C (red circles) and 36$^{\circ}$C (blue crosses) are essentially identical and show no frequency dependence as expected for the rheology of a colloidal gel in a low viscosity solvent. Scaling behavior of the smectic modulus for the 8CB+aerosil indicates that the enhanced elasticity and glassy rheology result from a dense network of screw dislocations that arise due to the quenched disorder and whose motion is pinned by the disorder~\cite{ranjini_8CB}. An excellent overlap of the data for
different disorder strengths is observed when the ratio $\frac{G^{\prime}}{{(x-1)}{G^{\prime\prime}}}$ is plotted {\em vs.} temperature below the pseudocritical region \cite{ranjini_8CB}. These observations point to a one-to-one
correspondence between $\rho_{s}$ and $x$, where $x$ quantifies the noise temperature influencing the dynamics.
\begin{figure}[ht4!]
\centering
\includegraphics[width=12.cm]{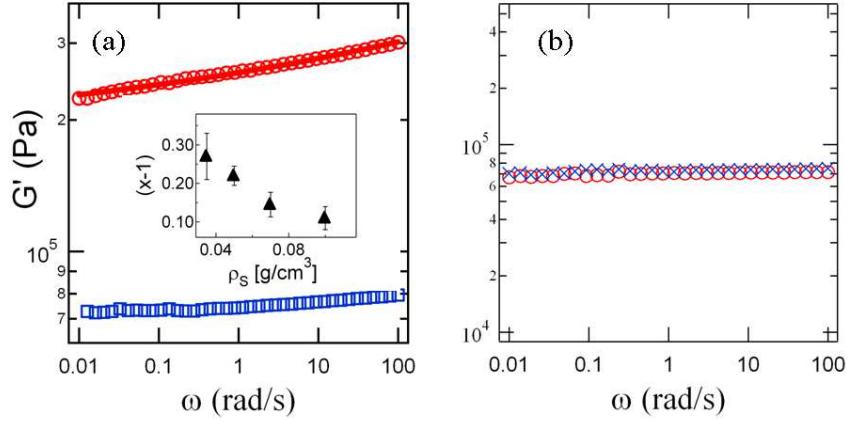}
\caption{\label{p1251} Rheology data for the elastic moduli $G^{\prime}$ of 8CB dispersed in a porous aerosil gel of density $\rho_{s}$ = 0.10 g/cm$^{3}$ is shown in a). The red circles correspond to G$^{\prime}$ measured in the smectic phase (T = 21$^{\circ}$C) which shows a power law contribution characterised by an exponent $(x-1)$. The solid line through the data at 21$^{\circ}$C is the result of a fit to a weak power law component plus a frequency-independent component as detailed in \cite{ranjini_8CB}. The viscous modulus $G^{\prime\prime}$ in the smectic phase also shows an identical power law dependence when the contribution of the gel is subtracted from the response (not shown in this figure). The blue squares correspond to the approximately frequency independent behaviour when 8CB is in the nematic phase (T = 36$^{\circ}$C). The plot of $(x-1)$ {\em vs.} $\rho_{s}$ is shown in the inset. The errors bars are computed from the temperature variation of $(x-1)$ at each $\rho_{s}$. This data is dapted from \cite{ranjini_8CB}. Part b) shows the frequency response of a simple organic liquid dibutyl phthalate confined to the pores of an aerosil gel ($\rho_{s} \sim$ 0.09 g/cm${^3}$). The $G^{\prime}$ measured at 21$^{\circ}$C (red circles) and 36$^{\circ}$C (blue crosses) are both frequency independent, while their magnitudes are almost identical to the G$^{\prime}$ measured for the 8CB-aerosil mixture (with $\rho_{s} \sim$ 0.10 gm/cm$^{3}$) when 8CB is in the nematic or isotropic phase.}  
\end{figure}

Microrheological measurements in living cells show power law frequency dependences of the viscoelastic moduli of the  
cytoskeleton, where the effective noise temperature lies in the range $1.1 < x < 1.3$ and is found to depend strongly on the nature of the biological intervention employed in the experiment (such as the addition of a contractile histamine) \cite{fabry_microrheol}. Those biological processes that cause contractile acivity in the cell or result in the polymerisation of the cell proteins can drive the system towards a glass transition. Disorder and metastability, therefore, contribute significantly to the mechanical functions of the living cell, with the noise temperature $x$ serving as a useful measure of its deformation and flow behaviour. Similar glassy behaviour is also observed in the rat airway smooth muscle during actin modulation \cite{rasm}. The scale-free soft glassy rheology of living cells has been a subject of recent research \cite{alcaraz,desprat,lenormand}. 

Under continuous shear flow, the potential energy landscape of soft glassy materials gets modified, such that the depths of the minima decrease temporally according to the relation $E(t) = E - \frac{1}{2}k{\dot\gamma}^{2}t^{2}$, where $k$ is the elastic constant relating the stress to the deformation and $\dot\gamma$ is the macroscopic strain rate. Many SGMs, such as concentrated emulsions  \cite{sriram_em} and foams \cite{doug_foam} show Herschel-Bulkley type flow curves (discussed in section 2.2) with a well-defined yield stress $\sigma_{y}$. For stresses of magnitude $\sigma > \sigma_{y}$, the flow curve assumes a power law form with a power ($x$-1) of around 0.1. Another example is a microgel paste made of jammed polyelectrolyte particles that can flow when sheared above a certain yield stress value. The glassiness of this material is evident from its Herschel-Bulkley flow curve \cite{cloitre_paste2}. The synthetic clay laponite \cite{lapo_sgr} also shows a power law flow curve with noise temperature $x$ = 1.20, a value slightly inconsistent with the values of $x$ obtained from the frequency-dependent measurements of $G^{\prime}$ and $G^{\prime\prime}$. This discrepancy leads the authors to conclude that the nonlinear flow behaviour of colloidal laponite is significantly more complex than the predictions of the SGR theory \cite{sollich_prl,sollich_sgr}. There is, therefore, a need for more experimental studies to verify the predictions of the SGR theory for the nonlinear response of SGMs.

\section{Concluding remarks}
\label{conclusion}

In this paper, we have reviewed theoretical models used extensively to understand the slow dynamics of soft and living matter, focussing in particular on two key concepts:  jamming and soft glassy rheology. The main purpose of the paper has been to provide a overview of these concepts and to catalogue some of the extensive experimental studies on a diverse range of soft and glassy materials that relate to these models. These concepts continue to undergo  
theoretical development, as exemplified by the recent work relating the zero-temperature jamming transition to k-core percolation \cite{schwarz_percolation} and the introduction of tensorial constitutive models that generalise SGR \cite{tensor_sgr}.  On the experimental side, the emergence of techniques such as multispeckle correlation spectroscopy and the availability of high-precision commercial rheometers have greatly facilitated the study of soft glassy materials.  Simultaneous rheometry and PCS studies that can correlate the bulk flow behaviour of these materials to their microscopic structure and dynamics are a potentially fruitful avenue for future experimental work.  Such  
continued innovation in experimental methods should lead to a better understanding of the aging, slow relaxational behavior, and glassy rheology observed in soft and living systems.

The authors gratefully acknowledge R. Colby, H. Yardimci, D. Sessoms, M. Borthwick, and S. Mochrie for their significant contributions to the
studies described in \cite{ranjini_lapo} and \cite{ranjini_8CB}.  They also thank S. Narayanan and A. Sandy for their expert assistance with the XPCS
studies.  RL acknowledges support of the NSF (Grant No. DMR-0134377).  Use
of the APS was supported by the DOE, Office of Basic Energy Sciences,
under Contract No.~W-31-109-Eng-38.


\end{document}